\documentclass[aps,prb,superscriptaddress,showkeys]{revtex4-2}
\usepackage[utf8]{inputenc}
\usepackage{amsmath,amssymb,amsfonts}
\usepackage{multirow}
\usepackage{graphicx,float}
\usepackage[colorlinks=true,
citecolor=blue,
linkcolor=blue,
urlcolor=blue]{hyperref}
\usepackage{footmisc}
\usepackage{array}
\newcolumntype{P}[1]{>{\centering\arraybackslash}p{#1}}

\begin{document}

\title{Hydrostatic Pressure Induced Anomalous Enhancement in the Thermoelectric Performance of Monolayer MoS$_{2}$}
\author{Saumen Chaudhuri}
\affiliation{Department of Physics, Indian Institute of Technology Kharagpur, Kharagpur 721302, India}
\author{Amrita Bhattacharya}
\affiliation{Department of Metallurgical Engineering and Materials Science, Indian Institute of Technology Bombay, Mumbai 400076, India}
\author{{A. K. Das}}
\affiliation{Department of Physics, Indian Institute of Technology Kharagpur, Kharagpur 721302, India}
\author{{G. P. Das}}
\affiliation{Research Institute for Sustainable Energy, TCG Centres for Research and Education in Science and Technology, Sector V, Salt Lake, Kolkata 700091, India}
\author{B. N. Dev}
\email[corresponding author: ]{bhupen.dev@gmail.com}
\affiliation{Department of Physics and School of Nano Science and Technology, Indian Institute of Technology Kharagpur, Kharagpur, 721302, India}
\affiliation{Centre for Quantum Engineering, Research and Education, TCG Centres for Research and Education in Science and Technology, Sector V, Salt Lake, Kolkata 700091, India}

\begin{abstract}
The hydrostatic pressure induced changes in the transport properties of monolayer (ML) MoS$_2$ have been investigated using first-principles density functional theory based calculations.  The application of pressure induces shift in the conduction band minimum (CBM) from K to $\Lambda$, while retaining the band extrema at K in around the same energy at a pressure of 10 GPa. This increase in valley degeneracy is found to have a significant impact on the electronic transport properties of ML-MoS$_2$ via enhancement of the thermopower (S) by up to 140\% and power factor (S$^{2}$$\sigma$/$\tau$) by up to 310\% at 300 K.  Besides, the very low deformation potential (E$_\text{DP}$) associated with the CB-$\Lambda$ valley results in a remarkably high electronic mobility ($\mu$) and relaxation time ($\tau$). Additionally, the application of pressure reduces the room temperature lattice thermal conductivity ($\kappa_\text{L}$) by 20\% of its unstrained value, owing to the increased anharmonicity and resulting increase in the intrinsic phonon scattering rates. The hydrostatic pressure induced increase in power factor (S$^{2}$$\sigma$) and the decrease in $\kappa_\text{L}$ act in unison to result in a substantial improvement in the overall thermoelectric performance (zT) of ML-MoS$_2$. At 900 K with an external pressure of 25 GPa, zT values of 1.63 and 1.21 are obtained for electron and hole doping, respectively, which are significantly higher compared to the zT values at zero pressure. For the implementation in a thermoelectric module where both n-type and p-type legs should be preferably made of the same material, the concomitant increase in zT of ML-MoS$_2$ for both types of doping with hydrostatic pressure can be highly beneficial.  
\end{abstract}

\keywords{DFT, hydrostatic pressure, MoS$_{2}$, thermoelectric properties}

\date{\today}
\maketitle

\section{Introduction}
The search for renewable energy sources has become a quintessential challenge for mankind \cite{owusu2016review}. Thermoelectric materials (TEMs), which can convert a difference in temperature into useful voltage based on the principle of Seebeck effect \cite{zhang2015thermoelectric, freer2020realising}, may aid to it via waste heat recovery \cite{orr2017prospects}. The efficiency of a thermoelectric material  depends on a dimensionless parameter called the thermoelectric figure of merit (zT), which is expressed as $\text{zT}=  \frac{\text{S}^{2}\sigma}{\kappa} \text{T}$. zT is directly proportional to the product of square of Seebeck coefficient S and the electrical conductivity $\sigma$, while is inversely proportional to the total thermal conductivity, $\kappa = \kappa_\text{e}+\kappa_\text{L}$, which is the sum of the electrical ($\kappa_\text{e}$) and the lattice thermal conductivity ($\kappa_\text{L}$), at a given temperature T. Tuning the zT of a given TEM is an extremely tedious task, since the transport parameters i.e. the S, $\sigma$, and $\kappa_\text{e}$ are inter-dependent on each other, as they all depend on the electronic carrier concentration (n). Therefore, the two main routes that can be followed to maximize the zT are (a) engineering the electronic band structure to optimize the power factor \cite{bilc2015low, pei2011convergence} and (b) phonon engineering for lowering the $\kappa_\text{L}$ of the material \cite{dresselhaus2007new, zhang2022reduced, gautam2020enhanced, li2012vacancy, ding2015manipulating}. 

More recently, the two dimensional planar materials have garnered a lot of interest in thermoelectric research owing to their unique properties arising from quantum confinement effects. The excellent combination of electrical and thermal transport in these 2D materials leads to a significantly large thermoelectric zT \cite{bilc2015low, dresselhaus2007new}. Over the years, various 2D materials, viz. transition metal dichalcogenides (TMDCs) (e.g., MoS$_2$, WS$_2$, MoSe$_2$, and HfS$_2$), group IVA-VA compounds (e.g., SnSe, SnS, GeSe) etc., have emerged as potential TEM with reasonably good thermoelectric performance \cite{bera2019strain, jin2015revisit, jia2022high, nguyen2018review}. In recent times, especially the layered TMDCs have drawn significant interest owing to their tunable band gap, high electrical mobility and low thermal conductivity \cite{wang2012electronics, guo2013theoretical, akinwande2014two, chaudhuri2023strain, radisavljevic2011single}. Some of the TMDCs, such as WS$_2$ \cite{patel2020high}, WSe$_2$ \cite{kumar2015thermoelectric}, HfS$_2$ \cite{wang2021improved}, SnSe$_2$ \cite{su2013snse2} etc., have shown notably high thermoelectric efficiency. Among the family of TMDCs, MoS$_2$ has been explored extensively as a potential thermoelectric material, both theoretically \cite{bhattacharyya2014effect, huang2014theoretical, wickramaratne2014electronic, jena2017compressive, guo2013high, xiang2019monolayer, chaudhuri2023strain, chaudhuri2023strain1} and experimentally \cite{hippalgaonkar2017high, kayyalha2016gate}, and the monolayer (ML) counterpart appears to be a reasonably good thermoelectric material \cite{bhattacharyya2014effect, jena2017compressive, xiang2019monolayer, hippalgaonkar2017high, kayyalha2016gate}. Kedar \textit{et al} \cite{hippalgaonkar2017high} have measured the room temperature thermoelectric power factor of ML-MoS$_2$ to be 8.5 mWK$^{-2}$m$^{-1}$, which is comparable to the well known thermoelectric materials, such as Bi$_2$Te$_3$. Buscema \textit{et al} \cite{buscema2013large} have found a large Seebeck coefficient in ML-MoS$_2$, which is also tunable by an external electric field. However, the zT value obtained with ML-MoS$_2$ is very low (i.e., a zT value of only $\sim 0.11$ is achieved with ML-MoS$_2$ at 500 K \cite{jin2015revisit}), which is not useful for practical purposes. The primary reason behind the low thermoelectric efficiency of ML-MoS$_2$ is derived from its rather high lattice thermal conductivity ($\kappa_\text{L}$), which is considerably higher compared to the other analogous TMDCs, such as WSe$_2$, ZrS$_2$, HfS$_2$ etc. \cite{bera2019strain, patel2020high, rai2020electronic, yan2014thermal, guo2019predicted}. Therefore, finding out a way by which both the power factor (S$^{2}$$\sigma$) and the $\kappa_\text{L}$ can be optimized simultaneously in a way to maximize the thermoelectric zT of ML-MoS$_2$ has been an extensive quest.

Efforts have been made to improve the thermoelectric efficiency of ML-MoS$_2$ by applying an external electric field \cite{buscema2013large} or mechanical strain \cite{ding2015manipulating, wang2021improved, bhattacharyya2014effect, jena2017compressive, xiang2019monolayer, chaudhuri2023strain1}. Doping with impurity atoms and hybridization are the widely explored avenues in improving the zT of a material \cite{gautam2020enhanced, kong2018realizing, gangwar2019ultrahigh}. In-plane strain engineering has also been employed routinely to tune the electronic and thermoelectric properties of various 2D materials, such as WS$_2$, ZrS$_2$, HfS$_2$, ZnO etc. \cite{bera2019strain, wang2021improved, lv2016strain, chaudhuri2023ab}. In one of our earlier studies, we have seen that in-plane tensile strains, in general, and strains along the zig-zag direction, in particular, can significantly enhance the thermoelectric efficiency of ML-MoS$_2$ \cite{chaudhuri2023strain1}. However, the implementation of different in-plane strains in experiments is a daunting task \cite{pu2013fabrication, castellanos2013local, bertolazzi2011stretching, jiang2014buckling, li2019molecular}. Application of hydrostatic pressure, on the other hand, can be a potential alternative easier route that can be achieved experimentally owing to the fairly simple procedure and reversibility. Previous theoretical and experimental studies on the pressure dependence of electronic properties, structural parameters and elastic constants of bulk MoS$_2$ suggest that hydrostatic pressure might be effective in tuning its transport properties \cite{guo2013high, peelaers2014elastic, fu2017k, chi2014pressure}. Zhang \textit{et al} demonstrated that the thermoelectric performance of bulk MoS$_2$ can be enhanced by the application of hydrostatic pressure \cite{guo2013high}. For ML-MoS$_2$, the variation in the thermoelectric parameters with various in-plane strain has been studied thoroughly over the years \cite{ding2015manipulating, wang2021improved, bhattacharyya2014effect, jena2017compressive, xiang2019monolayer, chaudhuri2023strain1}. However, studies on the effect of hydrostatic pressure on the transport properties of ML-MoS$_2$ are lacking.  

Interestingly, in a recent study, a direct K-K to indirect K-$\Lambda$ band gap transition has been optically detected in ML-MoS$_2$ under hydrostatic pressure \cite{fu2017k}. Therefore, a pressure induced shift of the conduction band minimum (CBM) from K to $\Lambda$ is inevitable. At the critical pressure, exploiting the degeneracy of the conduction band edges at K and $\Lambda$, an enhancement in the thermoelectric performance of ML-MoS$_2$ can be achieved. Also, the overall transport mechanism, both electronic and lattice, in ML-MoS$_2$, can be tuned continuously and controllably by the application of pressure. The hydrostatic pressure induced modification in the transport properties of ML-MoS$_2$ has not been studied theoretically or experimentally. Therefore, a thorough investigation of the underlying mechanism is necessary. This may pave the way for the application of ML-MoS$_2$ or any other semiconducting TMDCs in future commercial thermoelectric devices. In the present work, we have undertaken a theoretical investigation of hydrostatic pressure-mediated modification in the thermoelectric properties of ML-MoS2.

\section{Computational details}
First-principles calculations have been performed under the framework of ab-initio density functional theory (DFT) as implemented in the Vienna Ab Initio Simulation Package (VASP) \cite{kresse1996efficient, kresse1996efficiency} with projector augmented wave (PAW) potentials to account for the electron-ion interactions \cite{kresse1999ultrasoft}. The electronic exchange and correlation (XC) interactions are addressed within the generalized gradient approximation (GGA) of Perdew-Burke-Ernzerhof (PBE) \cite{perdew1996generalized}. In all calculations, the Brillouin zone (BZ) is sampled using a well-converged Monkhorst-Pack \cite{monkhorst1976special} k-point set ($21 \times 21 \times 1$), and a conjugate gradient scheme is employed to optimize the geometries until the forces on each atom are found to be less than 0.01 eV/\AA. A vacuum thickness of approximately 20 \AA \, has been used to avoid the spurious interaction between the periodic images of the layers. In order to generate the structure under hydrostatic pressure, the bulk lattice of MoS$_2$ is subjected to external stress of equivalent amount. The electronically optimized strained structure of bulk MoS$_2$  is then used to cleave one single layer of the MoS$_2$ from its bulk lattice under different hydrostatic pressures i.e. for 5, 10 and 25 GPa. A similar approach has also been adopted to generate hydrostatic strained structures in earlier reports \cite{fu2017k, fan2015electronic}. 

The temperature- and carrier concentration-dependent changes in the thermoelectric parameters such as the Seebeck coefficient (S), electrical conductivity ($\sigma$), power factor (S$^{2}\sigma$) etc. have been calculated by using the energy eigen values to solve the semi-classical Boltzmann transport equation as implemented within the BoltzTraP code \cite{madsen2006boltztrap}. All the transport properties have been calculated within the constant relaxation time approximation (CRTA) which assumes that the charge carrier relaxation time does not vary with energy or carrier concentration. To go beyond the CRTA, the charge carrier mobility ($\mu$) and relaxation time ($\tau$) are explicitly determined using the acoustic deformation potential (ADP) theory developed by Bardeen and Shockley \cite{bardeen1950deformation}. Considering the scattering between charge carriers and longitudinal acoustic phonons, the $\mu$ and $\tau$ of the 2D materials are computed as $\mu = \frac{2e{\hbar}^{3}\text{C}_\text{2D}}{3{\text{k}_\text{B}}\text{T}{\text{m}^{*}}^{2}(\text{E}_\text{DP})^{2}}$ and $\mu = \frac{e\tau}{\text{m}^{*}}$, where $\text{C}_\text{2D}$ is the effective elastic constant, $\text{m}^*$ is the carrier effective mass, and $\text{E}_\text{DP}$ is the deformation potential constant corresponding to the conduction and the valence band edges. The material-specific inputs such as the $\text{C}_\text{2D}$, $\text{m}^*$, and the $\text{E}_\text{DP}$ are determined from first-principles calculations using VASP. 

The phonon dispersion curves are calculated based on the supercell approach using the finite displacement method restricting the atomic vibrations to an amplitude of 0.015 \AA \, as implemented in the phonopy code \cite{togo2015first}. To compute the lattice transport properties, the Boltzmann Transport Equation (BTE) for phonons is solved under the relaxation time approximation (RTA) as implemented in the phono3py code \cite{togo2015distributions}. The second- and third-order interatomic force constants (IFC) are calculated using convergence-checked $4 \times 4 \times 1$ and $2 \times 2 \times 1$ supercells based on the relaxed unit cell, respectively. Fourth- and higher-order IFCs are not taken into consideration due to their presumably small contribution to lattice thermal transport. Forces are optimized using a fine energy convergence criterion of 10$^{-8}$ eV. Well-converged k-meshes are used to optimize the structures. Supercell method is used to ensure convergence, whereby convergence is reached for $4 \times 4 \times 1$ supercell. To accurately compute the lattice thermal conductivity, a dense q-mesh of $51 \times 51 \times 1$ is used to sample the reciprocal space of the primitive cells. The mode-resolved phonon transport parameters, such as the phonon group velocity (v$_\lambda$) and relaxation time ($\tau_\lambda$) are extracted using python-based extensions.

\begin{figure}[h!]
	\centering
	\includegraphics[scale=0.35]{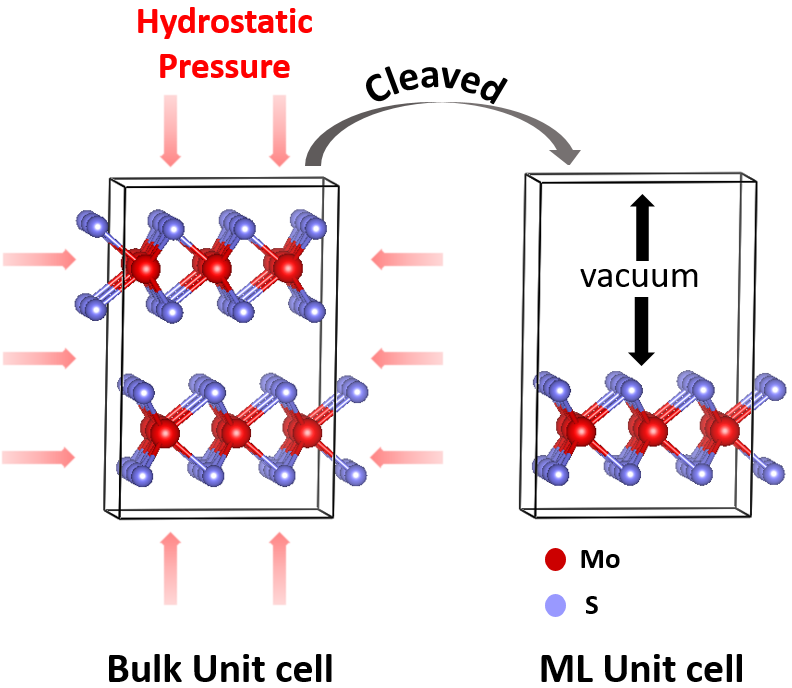}
	\caption{ Schematic diagram illustrating the procedure of obtaining a hydrostatic strained MoS$_2$ monolayer by cleaving its equivalent strained bulk MoS$_2$ counterpart.}
	\label{Fig.1}
\end{figure}

\section{Results and Discussion}
\subsection{Structure parameters and electronic properties}

The single-layer counterpart of MoS$_{2}$ forms a hexagonal honeycomb crystal structure with the Mo and S atoms arranged in a trigonal prismatic coordination, as can be seen from Fig. \ref{Fig.1}. A monolayer (ML) structure of MoS$_{2}$ consists of one Mo and two S atomic planes arranged in a “sandwich” (S-Mo-S) type structure. The primitive unit cell of bulk MoS$_{2}$ comprises of two such monolayers with an inter-layer separation (t) of 6.15 \AA. The structural parameters of both the bulk- and the ML-MoS$_{2}$, calculated herein, are in good agreement with earlier reports \cite{jena2017compressive, chaudhuri2023strain1}. To study the effect of hydrostatic pressure on the electronic and transport properties of ML-MoS$_{2}$, different pressures of up to 25 GPa have been applied. A schematic of the pressure application on ML-MoS$_{2}$ is presented in Fig. \ref{Fig.1}. To confirm the dynamic stability of the crystal structure, phonon dispersion has been calculated at each pressure. From the electronic structure calculations, ML-MoS$_{2}$ in its pristine form is found to be a direct band gap semiconductor with both the valence band maximum (VBM) and the conduction band minimum (CBM) located at the same high-symmetry point i.e. at the K (0.33, 0.33, 0) point of the Brillouin zone (BZ). The band gap value of pristine ML-MoS$_{2}$ is found to be 1.68 eV, which is in agreement with previous theoretical works \cite{chaudhuri2023strain, jena2017compressive, das2014microscopic}.

\begin{figure}[h!]
	\centering
	\includegraphics[scale=0.4]{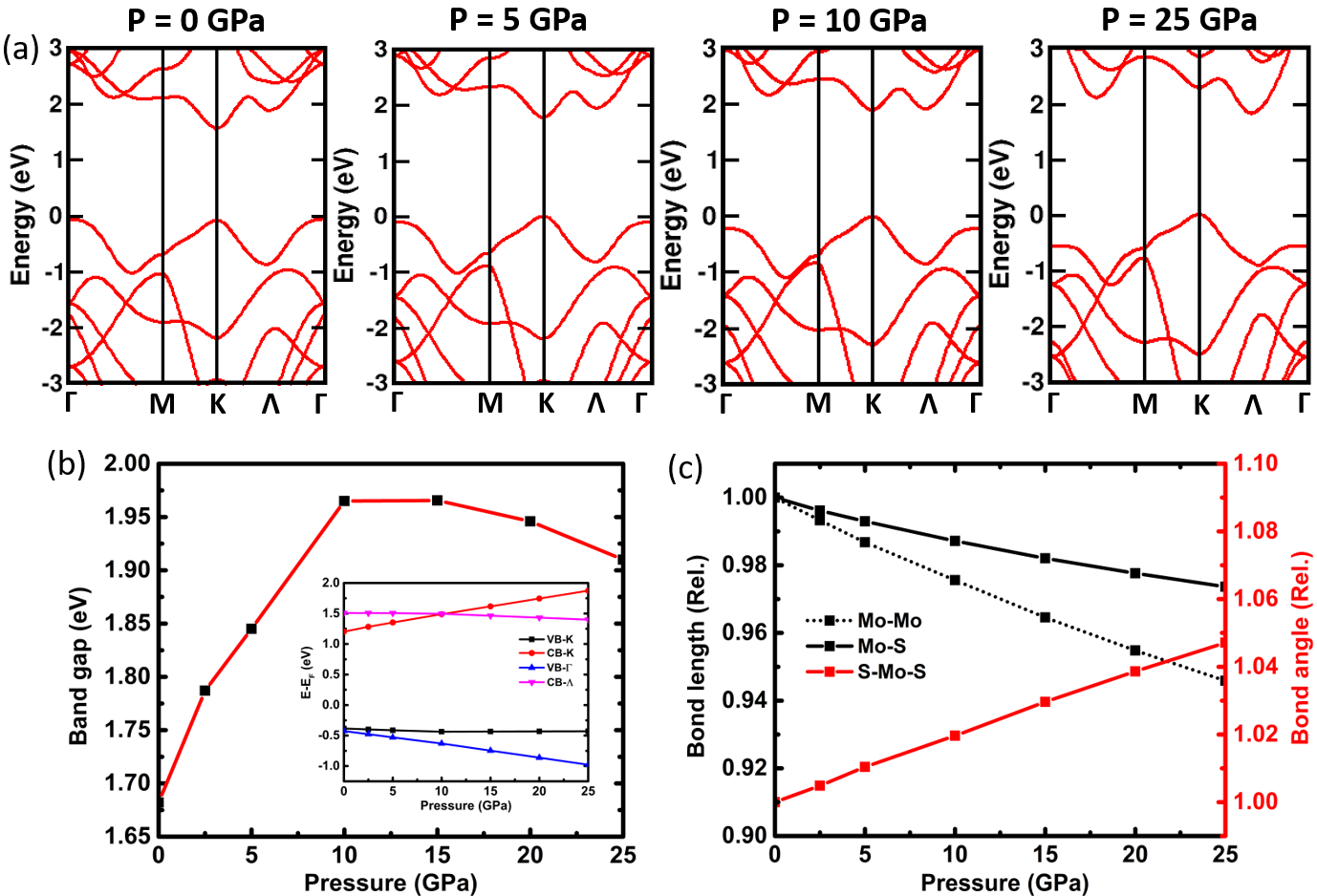}
	\caption{ (a) Band dispersion along the high symmetry path of ML-MoS$_2$ at different values viz. 0 GPa, 5 GPa, 10 GPa, and 25 GPa of hydrostatic pressure. The pressure values are given at
		the top of the corresponding band structure plots. (b) The variation in the band gap as a function of the hydrostatic pressure (b: inset shows the energies corresponding to the
		valence and conduction band edges at different k-points with reference to the Fermi energy E$_\text{F}$). (c) The various structural parameters i.e. the bong length (shown in black) and bond angle (shown in red) of ML-MoS$_2$ as a function of hydrostatic pressure.}
	\label{Fig.2}
\end{figure} 

The hydrostatic pressure induced effects on the structural parameters and electronic structure of ML-MoS$_2$ are presented in Fig. \ref{Fig.2}. From the band structure plots it can be seen that, with the application of pressure, the CBM at K point (CB-K) shifts toward higher energy (i.e. away from the Fermi level), while the energy shift in the valence band top at K (VB-K) is nominal. Therefore, the band gap increases initially, until P = 10 GPa, as can be seen from the variation in the band gap with pressure shown in Fig. \ref{Fig.2} (b). At a pressure around 10 GPa, the CBM shifts to the $\Lambda$-point  i.e., midway along the K-$\Gamma$ path, while the one extrema at the K point lying very close in energy, thereby forming degenerate energy valleys in the conduction band. Such valley or band degeneracies are found to have useful implications in enhancing the thermoelectric performance of a material \cite{bilc2015low, pei2011convergence}. Starting from 15 GPa, when the CBM is at the $\Lambda$ point, the band gap starts to decrease with increasing pressure. To quantify the energy shifts, the energies corresponding to the VB and the CB edges are plotted as a function of pressure and presented in Fig. \ref{Fig.2} (b). It is clear that the energy shifts in the CB-K and the VB-$\Gamma$ are significantly higher compared to the energy shifts in other valence and conduction band extrema. 

To understand these energy shifts in greater detail, we have calculated the band decomposed charge densities corresponding to the relevant valence and conduction band edges, such as the CB at K (CB-K), CB at $\Lambda$ (CB-$\Lambda$), VB at K (VB-K) and VB at $\Gamma$ (VB-$\Gamma$) (see Fig. S1 in supplementary material). The band edges at different k-points in the BZ are found to have very different real space charge densities, however, mostly comprising of the Mo-$d$ orbital. For example, the states at CB-K and VB-K are predominantly comprising of Mo-d character arising from d$_{z^{2}}$ and d$_{x^{2}-y^{2}}$ orbitals, respectively. The states at CB-$\Lambda$ and VB-$\Gamma$, on the other hand, have a certain admixture of S-$p$ orbitals along with the Mo-$d$ orbitals. With the application of pressure, both the Mo-Mo and Mo-S bond length decreases with an increase in the S-Mo-S bond angle and therefore, the S-S interplanar distance, as can be seen from Fig. \ref{Fig.2} (c). However, the magnitude of reduction in Mo-Mo bond length is much higher compared to that of Mo-S bond length. It is, therefore, expected that the states that are composed purely of Mo-$d$ orbitals (such as CB-K), and thus depend entirely on the Mo-Mo bond length, will experience larger energy shifts compared to the states that comprise of both Mo-$d$ and S-$p$ character (such as CB-$\Lambda$). The large energy shifts of the CB-K, which is the most relevant one in this case, can be understood by analysing the pressure induced modifications in the structural parameters. With increasing bond angle and decreasing bond length under pressure, the interaction strength between the Mo-$d_{x^{2}-y^{2}}$ and S-$p$ orbitals weakens, while the coupling between the $d_{z^{2}}$ orbitals will strengthen. Therefore, the energy splitting of the pair of states with predominantly Mo-$d_{z^{2}}$ character, such as at CB-K and VB-$\Gamma$, increases under pressure and thereby, shifting the CB-K towards the higher energy.

\subsection{Thermoelectric transport properties}

\begin{figure}[h!]
	\centering
	\includegraphics[scale=0.37]{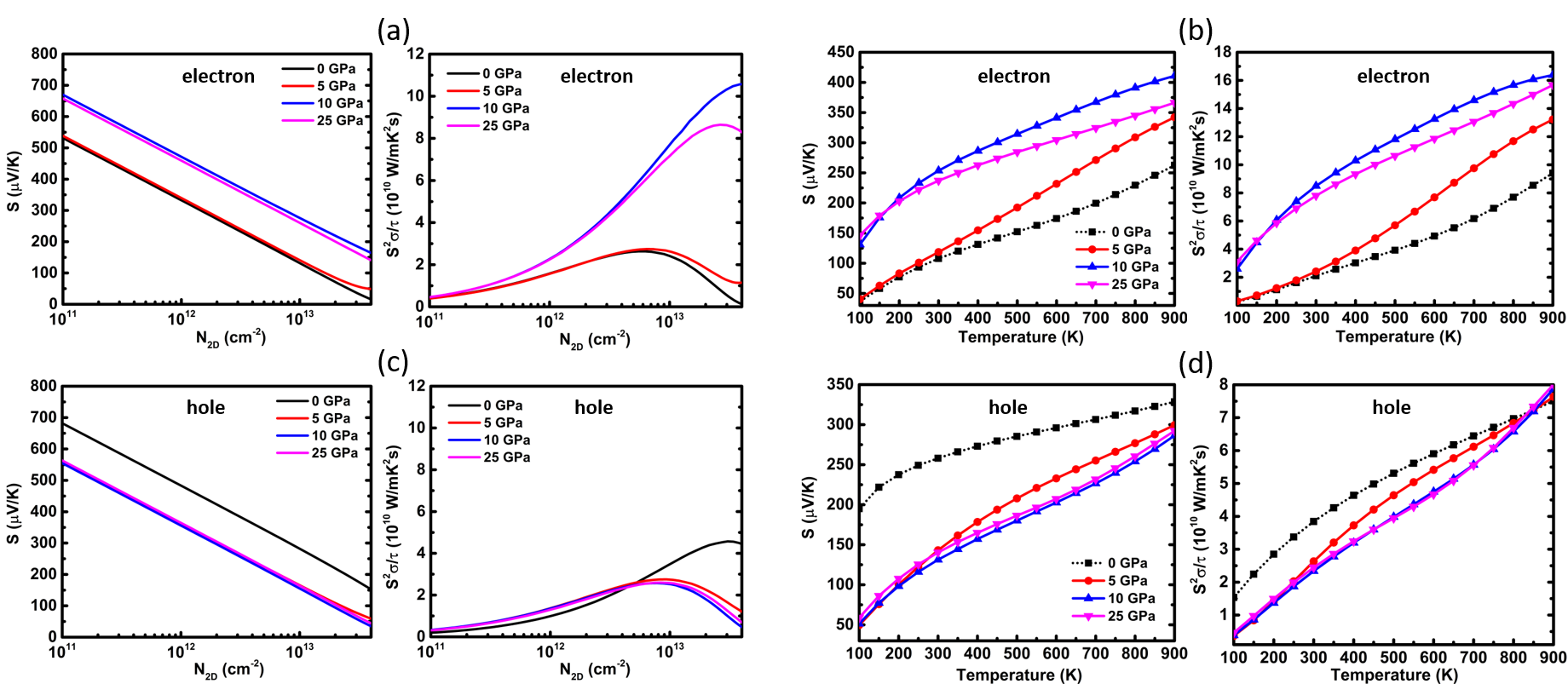}
	\caption{Variation in the thermoelectric parameters i.e., the thermopower (S) and relaxation time ($\tau$) scaled power factor (S$^{2}\sigma$/$\tau$) of ML-MoS$_2$ for different values of applied hydrostatic pressure as a function of (a, c) carrier concentration (N$_\text{2D}$) at a fixed temperature of 300 K and (b, d) as a function of temperature for a fixed electron (upper panel) and hole (lower panel) doping of $1.3 \times 10^{13}$ /cm$^{2}$.}
	\label{Fig.3}
\end{figure}

To understand the implications of the pressure induced modifications in the electronic band structure of ML-MoS$_{2}$, various thermoelectric transport parameters, such as the Seebeck coefficient or thermopower (S), electrical conductivity ($\sigma$), power factor (PF: S$^{2}\sigma$) and the charge carrier mobility ($\mu$) have been investigated as a function of increasing hydrostatic pressure. Specifically, the pressure induced degeneracy in the conduction band is expected to have a significant impact on the thermoelectric parameters of ML-MoS$_{2}$. The pressure induced variation in the S and PF as a function of carrier concentration at a fixed temperature T= 300 K, and as a function of temperature at a fixed carrier concentration of N$_\text{2D}$ = $1.3 \times 10^{13}$ /cm$^{2}$ are presented in Fig. \ref{Fig.3} for both the n-type (shown in top panel) and p-type (shown in the bottom panel) doping. With increasing pressure, the S is found to increase significantly for n-type doping concentration of $0.01-3 \times 10^{13}$ /cm$^{2}$. The highest value of S can be achieved at a hydrostatic pressure of 10 GPa, which is due to the pressure induced valley degeneracy at CB-K and CB-$\Lambda$. At this pressure owing to the increased valley degeneracy at the CB edge, the density of states effective mass (m${_\text{d}}^*$) increases, thereby resulting in an enhancement of the S, without explicitly affecting the charge carrier mobility ($\mu$). Therefore, the power factor is significantly improved due to the degenerate valleys. However, with pressure the valley degeneracy at the valence band extrema is lost, since the band maximum at the $\Gamma$ point shifts to lower energy and the band maximum is found to lie only at the K-point. Due to this loss in valley degeneracy in the valence band, a reduction in S is incurred upon hole doping.   For both electron and hole doping, the thermopower decreases with increasing carrier concentration, which is a typical characteristic of any semiconducting material. Owing to the increase in S, a significant enhancement in the relaxation time ($\tau$) scaled power factor (S$^{2}\sigma$/$\tau$) is observed as well under the application of hydrostatic pressure for electron doping (see Fig. \ref{Fig.3} (a) and (b)).  The power factor value reaches up to $8.5 \times 10^{10}$ W/mK$^{2}$s at 300 K for electron doping at a pressure of 10 GPa, which is significantly higher compared to the zero pressure power factor values. The power factor represents the ability of a material to produce useful electrical power at a given temperature gradient. Therefore, the large power factor is indicative of better thermoelectric performance. Also, the Goldsmid-Sharp relation \cite{goldsmid1999estimation} is found to hold well, as the variation in the maximum achievable thermopower (S$_\text{max}$) with applied pressure at a fixed temperature follows the same trend as that of the band gap (see Fig. S2 in supplementary material). The relation is given as, E$_\text{g}$ = 2eS$_\text{max}$T, where E$_\text{g}$ is the band gap, and S$_\text{max}$ is the maximum attainable thermopower at any temperature T. It is, therefore, clear that both the electronic and the thermoelectric parameters are equally sensitive to the applied pressure. 

In order to understand the temperature dependency, the variation in the thermoelectric parameters of ML-MoS$_{2}$ is investigated for temperature in a range of 100 K-900 K at a fixed carrier concentration of $1.3 \times 10^{13} /\text{cm}^{2}$, which corresponds to $5 \times 10^{19} /\text{cm}^{3}$ in bulk configuration. The upper limit of the temperature is well within the thermal stability regime of ML-MoS$_2$, which was analyzed using molecular dynamics (MD) calculations in our earlier work \cite{chaudhuri2023strain1}. The S and PF are plotted as function of temperature for different hydrostatic pressure under electron and hole doping (see Fig. \ref{Fig.3} (b) and (d)). For the electron doping, both these quantities are found to be highest at a pressure of 10 GPa for the entire temperature range due to the increased degeneracy at the CBM. However, for the hole doping the pristine one is found to show higher S and PF as compared to their pressurized counterparts for the entire temperature range, which is due to the loss of the valley degeneracy at the VBM under pressure (see Fig. \ref{Fig.3} (d)). For both electron and hole doping, the PF is found to increase dramatically with increasing temperature. At 900 K, the highest PF values obtained for electron and hole doping are $16 \times 10^{10}$ W/mK$^{2}$s and $8 \times 10^{10}$ W/mK$^{2}$s, respectively. This is indicative of the possible use of ML-MoS$_{2}$ as a high temperature thermoelectric material. 

The transport parameters calculated herein, such as the $\sigma$, $\kappa_{\text{e}}$ and PF (S$^{2}\sigma$) are scaled by the charge carrier relaxation time ($\tau$), since the calculations performed here are based on the constant relaxation time approximation (CRTA). CRTA assumes that the carrier relaxation time ($\tau$) and therefore, the mobility ($\mu$) does not vary strongly with energy. Such an oversimplified assumption often lead to misleading conclusions. In the present case, the explicit determination of $\tau$ and $\mu$ is essential, since $\tau$ and $\mu$ itself are functions of external applied pressure. In the calculations of $\mu$ and $\tau$, only the scattering of charge carriers with longitudinal acoustic phonons within the acoustic deformation potential (ADP) theory of Bardeen and Shockley \cite{bardeen1950deformation} is considered (see Table 1 in supplementary material). It is assumed that the scattering with the polar optical phonons will have an inconsequential effect owing to the intrinsically non-polar nature of MoS$_{2}$. Other scattering events, such as boundary scattering and impurity scattering are also ignored, since these are strongly dependent on experimental conditions. Considering only a part of the total scattering processes, the calculations performed herein are expected to provide only an intrinsic limit of $\mu$ and $\tau$, which can be largely overestimated compared to the experimental values. 

\begin{figure}[h!]
	\centering
	\includegraphics[scale=0.45]{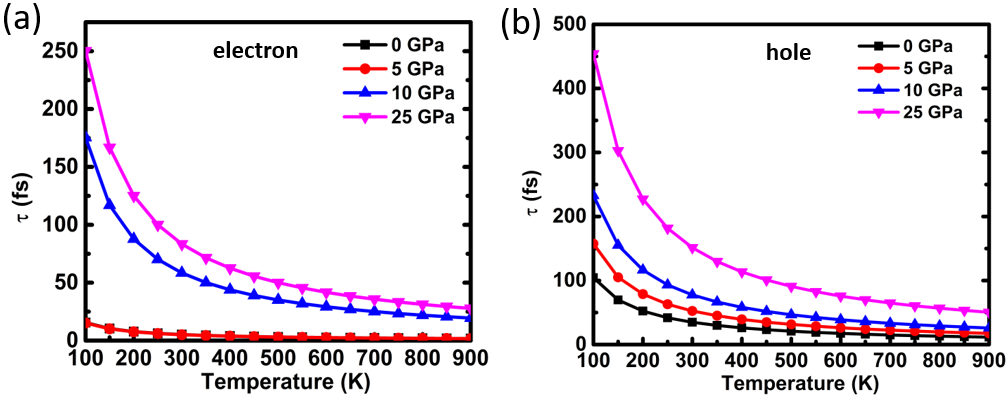}
	\caption{Variation in the charge carrier relaxation time ($\tau$) of ML-MoS$_{2}$ with temperature at different pressure values for (a) electron and (b) hole doping. }
	\label{Fig.4}
\end{figure}

\begin{table}
	\centering
	\renewcommand{\arraystretch}{1.5}
	\caption{\label{tab:Table 1} The calculated effective 2D elastic constant (C$_\text{2D}$), the carrier effective mass (m$^*$), the absolute deformation potential constant (E$_\text{DP}$) corresponding to the conduction and valence band edges and the total charge carrier mobility ($\mu$) of ML-MoS$_{2}$ at 300 K for different hydrostatic pressures. The m$^*$ and E$_\text{DP}$ are given for the different band edge positions (K and $\Lambda$ given in bracket).}
	\begin{tabular}{|P{2.5cm}|P{1.5cm}|P{1.5cm}|P{3.5cm}|P{3.5cm}|P{3.5cm}|}
		\hline 
		P (GPa) & Carriers & C$_\text{2D}$ & m$^*$ (m$_{0}$) & E$_\text{DP}$ (eV) & $\mu$ (cm$^{2}$V$^{-1}$s$^{-1}$)\tabularnewline
		\hline 
		\hline 
		\multirow{2}{*}{0} & e & \multirow{2}{*}{212.5} & $0.47$ (K) & $6.61$ (K) & 193.64\tabularnewline
		\cline{2-2} \cline{4-6} \cline{5-6} \cline{6-6} 
		& h &  & $0.58$ (K) & $2.29$ (K) & 1056.25 \tabularnewline
		\hline 
		\multirow{2}{*}{5} & e & \multirow{2}{*}{230.0} & $0.48$ (K) & $6.88$ (K) & 186.24 \tabularnewline
		\cline{2-2} \cline{4-6} \cline{5-6} \cline{6-6} 
		& h &  & $0.58$ (K) & $1.94$ (K) & 1597.22 \tabularnewline
		\hline 
		\multirow{2}{*}{10} & e & \multirow{2}{*}{246.2} & $0.51$ (K), $0.52$ $(\Lambda)$  & $6.97$ (K), $1.99$ $(\Lambda)$  & 1950.69 \tabularnewline
		\cline{2-2} \cline{4-6} \cline{5-6} \cline{6-6} 
		& h &  & $0.59$ (K) & $1.63$ (K) & 2308.22\tabularnewline
		\hline 
		\multirow{2}{*}{15} & e & \multirow{2}{*}{295.0} & $0.52$ $(\Lambda)$ & $1.43$ $(\Lambda)$ & 2782.38 \tabularnewline
		\cline{2-2} \cline{4-6} \cline{5-6} \cline{6-6} 
		& h &  & $0.64$ (K) & $1.22$ (K) & 4095.06\tabularnewline
		\hline 
	\end{tabular}
	\renewcommand{\arraystretch}{1.}
\end{table}

The charge carrier relaxation time ($\tau$) of ML-MoS$_2$ as a function of temperature at different pressure values is shown in Fig. \ref{Fig.4}, and the carrier mobility ($\mu$) plots are provided in the supplementary material (see supplementary material Fig. S3). The physical parameters that are required for the analytical formulation of the deformation potential theory (DP), such as the elastic constant (C$_\text{2D}$), carrier effective mass (m$^*$) and the deformation potential (E$_\text{DP}$) are presented in Table \ref{tab:Table 1}, and the computed mobility of ML-MoS$_2$ in the zero pressure case is in good agreement with previous reports \cite{wiktor2016absolute, hung2018two, rawat2018comprehensive}. Note that, due to the in-plane isotropic nature of ML-MoS$_{2}$, the values of the various physical parameters calculated along different directions are found to be identical. Therefore, the simplest form of the Bardeen and Shockley equation, which takes into account the average values of the C$_\text{2D}$, m$^*$, and E$_\text{DP}$ is implemented in this work. The mobility of electrons and holes at 300 K is found to be 193.6 and 1056.2 cm$^{2}$V$^{-1}$s$^{-1}$, respectively. The relaxation time corresponding to electron and hole scattering decreases with temperature following a parabolic function, as can be seen from Fig. \ref{Fig.4} (a) and (b). For both types of carriers, $\tau$ increases with increasing pressure. For electrons, $\tau$ increases anomalously with pressure, whereas for holes it increases in a monotonic fashion. A sharp increase in $\mu$ and $\tau$ starting from P = 10 GPa is seen for electron doping (see Fig. \ref{Fig.4} (a)). As the pressure increases to 10 GPa, the involvement of the conduction band valley at $\Lambda$ (CB-$\Lambda$) in electronic conduction becomes apparent. It can be seen from Table \ref{tab:Table 1} that, the effective mass of electrons is nearly identical in both the conduction band valleys at K and $\Lambda$. However, the deformation potential (E$_\text{DP}$) associated with the CB-$\Lambda$ is more than three times smaller than that with the CB-K. Therefore, the electronic mobility and relaxation time at the CB-$\Lambda$ valley is significantly higher compared to that at the CB-K. Thus, the contribution from the CB-$\Lambda$ results in the remarkable increase in total $\mu$ and $\tau$. For holes, on the other hand, the effective mass increases slightly With increasing pressure, however, due to the large reduction in E$_\text{DP}$ corresponding to the VBM at K, the hole mobility and therefore, the relaxation time increases. The enhancement in the mobility ($\mu$) and relaxation time ($\tau$) for both electrons and holes is highly beneficial for better thermoelectric performance. The estimated $\tau$ values are incorporated in the calculated electronic transport parameters, such as the $\sigma/\tau$, $\kappa_\text{e}/\tau$ and PF (S$^{2}\sigma/\tau$) to calculate the figure of merit (zT) as discussed in the subsequent paragraph. 

\begin{figure}[h!]
	\centering
	\includegraphics[scale=0.6]{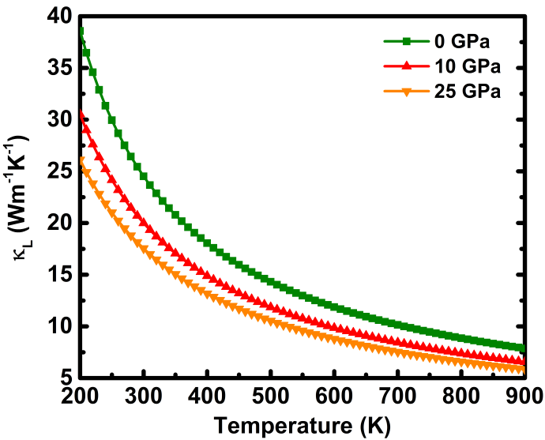}
	\caption{Variation in the lattice thermal conductivity ($\kappa_{\text{L}}$) of ML-MoS$_{2}$ with temperature at different values of hydrostatic pressure. }
	\label{Fig.5}
\end{figure}

The pressure induced modifications in the structural parameters are expected to have certain implications on the lattice transport properties. Therefore, the effect of pressure on the lattice thermal conductivity ($\kappa_{\text{L}}$) of ML-MoS$_2$ has been calculated and the variation in $\kappa_{\text{L}}$ with temperature at some representative values of pressure is shown in Fig. \ref{Fig.5}. For pristine ML-MoS$_{2}$, the $\kappa_{\text{L}}$ at 300K is found to be 24.41 Wm$^{-1}$K$^{-1}$, which agrees well with earlier theoretical \cite{rai2020electronic, cai2014lattice} and experimental reports \cite{yan2014thermal}. The $\kappa_{\text{L}}$ of ML-MoS$_{2}$ decreases with increasing pressure and reduces to 17.48 Wm$^{-1}$K$^{-1}$ at a pressure of 25 GPa. Owing to the inverse relationship of $\kappa_{\text{L}}$ with the thermoelectric figure of merit (zT), a reduction in $\kappa_{\text{L}}$ is immensely advantageous in enhancing the thermoelectric performance. The variation in $\kappa_{\text{L}}$ with temperature is found to follow the 1/T law owing to the increased probability of Umklapp scattering. A reduced value of 5.77 Wm$^{-1}$K$^{-1}$ is obtained at 900 K under an external pressure of 25 GPa. The total thermal conductivity ($\kappa$), which is a sum of the electronic ($\kappa_{\text{e}}$) and the lattice contribution ($\kappa_{\text{L}}$), is found to be largely dominated by the lattice counterpart, at all temperature and pressure. This can be explained as the general behaviour of a semiconductor that majority of the heat is carried by the phonons with negligible contribution stemming from the electrons. 

\begin{figure}[h!]
	\centering
	\includegraphics[scale=0.5]{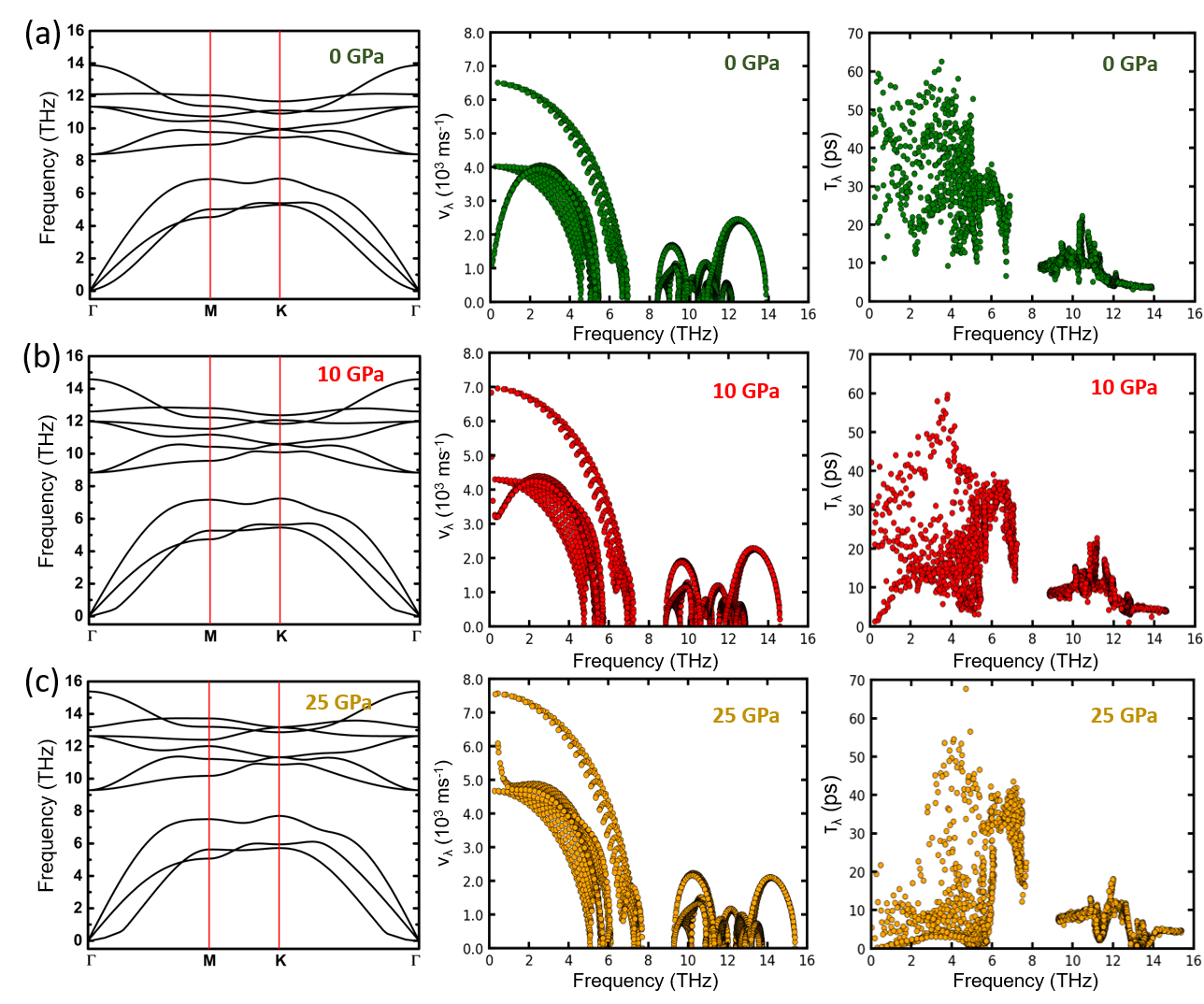}
	\caption{Phonon band structure (left) along the high-symmetry path $\Gamma$-M-K-$\Gamma$, (middle) group velocity (v$_\lambda$), and (right) scattering time ($\tau_{\lambda}$) of ML-MoS$_{2}$ at 300 K and at different hydrostatic pressure of (a) 0 GPa, (b) 10 GPa, and (c) 25 GPa. }
	\label{Fig.6}
\end{figure}

To investigate the pressure induced reduction in $\kappa_{\text{L}}$ in greater detail, further analysis of the $\kappa_{\text{L}}$ is performed. The mode-resolved values of the phonon group velocity (v$_\lambda$) and scattering time ($\tau_{\lambda}$) for a particular phonon mode $\lambda$ are computed at 300 K as a function of phonon frequency at different pressures and presented with the corresponding phonon band structures in Fig. \ref{Fig.6}. It can be seen that, both v$_\lambda$ and $\tau_{\lambda}$ of ML-MoS$_{2}$ is significantly higher within the frequency range of 0 to 7 THz, which corresponds to the acoustic phonon modes. It is, therefore, clear that the $\kappa_{\text{L}}$ stemming from the acoustic modes is much higher than the optical modes. With increasing pressure, due to the strengthening of the interatomic bonds, a stiffening of the acoustic phonon modes and a blue shift in the frequency of the optical phonon modes at the zone centre (at $\Gamma$) is observed. Thereby, the v$_\lambda$ associated with the acoustic phonon modes increases slightly. However, the phonon scattering time ($\tau_{\lambda}$) reduces significantly with increasing pressure, as can be seen in Fig. \ref{Fig.6}. Owing to the large decrease in $\tau_{\lambda}$, the $\kappa_{\text{L}}$ decreases with pressure.

\begin{figure}[h!]
	\centering
	\includegraphics[scale=0.4]{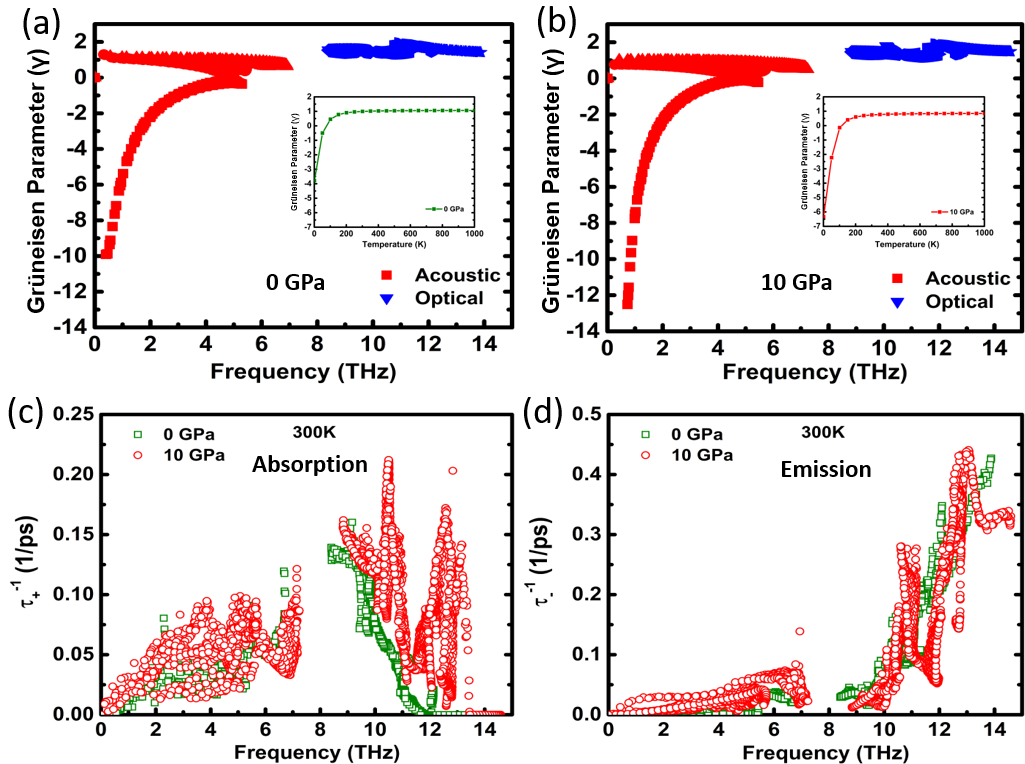}
	\caption{Calculated mode Gr$\ddot{\text{u}}$neisen parameters ($\gamma$) of ML-MoS$_{2}$ as a function of phonon frequency at (a) 0 GPa and (b) 10 GPa. The variation in the Gr$\ddot{\text{u}}$neisen parameter with temperature at 0 and 10 GPa is shown in the insets of (a) and (b). The phonon scattering rates at 300 K under 0 and 10 GPa hydrostatic pressure corresponding to the (c) absorption and (d) emission processes. }
	\label{Fig.7}
\end{figure}

\begin{figure}[h!]
	\centering
	\includegraphics[scale=0.4]{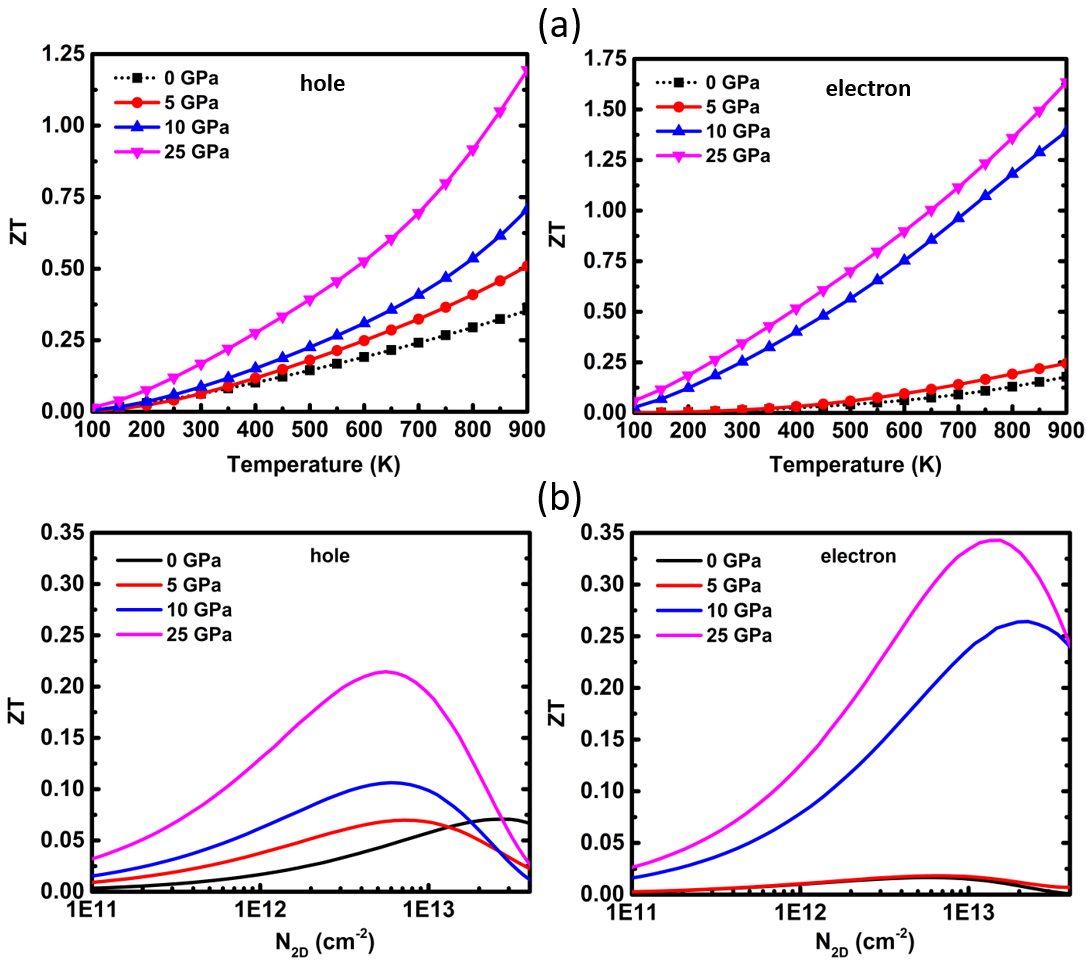}
	\caption{Variation in the thermoelectric figure of merit (zT) of ML-MoS$_{2}$ under hole and electron doping at various hydrostatic pressure as a function of (a) temperature at a fixed carrier concentration of $1.3 \times 10^{13}$ /cm$^{2}$ and (b) carrier concentration at a fixed temperature of 300 K. }
	\label{Fig.8}
\end{figure}

In order to understand the pressure induced large reduction in $\tau_{\lambda}$, the crystal anharmonicity and three-phonon scattering rate of ML-MoS$_{2}$ are computed at 300 K for two representative pressures, 0 GPa and 10 GPa, and shown in Fig. \ref{Fig.7}. The phonon scattering rate is strongly influenced by the crystal anharmonicity. In fact, it is the anharmonic vibration that limits the $\kappa_{\text{L}}$ to a finite value. The bond anharmonicity measures the asymmetry in atomic vibration, which can be characterized by the mode Gr$\ddot{\text{u}}$neisen parameters ($\gamma_\text{i}$), defined as $\gamma_\text{i}$ = -$\frac{\text{V}_0}{\omega_\text{i}}$ ${\frac{\Delta \omega_\text{i}}{\Delta \text{V}}}$, where $\omega_\text{i}$ is the frequency of the phonon mode i at the equilibrium volume V$_0$. The Gr$\ddot{\text{u}}$neisen parameter ($\gamma$) of ML-MoS$_{2}$ has the highest magnitude in the low frequency regime (0-2 THz), which belongs to the low frequency acoustic modes, albeit mainly the out-of-plane acoustic mode or the ZA mode. With increasing pressure, the mode Gr$\ddot{\text{u}}$neisen parameter ($\gamma_\text{i}$) corresponding to the ZA mode increases, as can be seen from Fig. \ref{Fig.7} (a) and (b). Therefore, the anharmonic phonon scattering rates are expected to increase with increasing pressure. Now, the anharmonicity giving rise to the scattering events involving three phonons has two possible scattering channels. One is the absorption process (+), where two phonons combine to produce the third phonon. The other one is the emission process (-), where one phonon splits into two different phonons. The scattering rates corresponding to the two processes, absorption (${\tau_{+}}^{-1}$) and emission (${\tau_{-}}^{-1}$), are plotted as a function of phonon frequency (see Fig. \ref{Fig.7}), and the total scattering rate (${\tau}^{-1}$= ${\tau_{+}}^{-1}$ + ${\tau_{-}}^{-1}$) is provided in the supplementary material (see Fig. S4 in supplementary material). Due to the energy conservation restriction, the absorption process (+) is dominant in the low frequency regime, whereas the emission process (-) is dominant in the high frequency regime. With increasing pressure, the scattering rates associated with both processes increase, mainly the absorption one (${\tau_{+}}^{-1}$), resulting in the decrease of the phonon relaxation time. The increase in the probability of phonon scattering is attributed to the shortening of the interatomic distances and the increased bond anharmonicity under the application of pressure.

Finally, combining all the calculated transport parameters, such as the S, S$^{2}\sigma$, $\tau$, $\kappa$, the thermoelectric efficiency or the zT of ML-MoS$_{2}$ has been estimated and its variation with temperature (T) and carrier concentration (N$_\text{2D}$) has been explored under different pressure conditions (see Fig. \ref{Fig.8}). The room temperature zT values obtained with pristine ML-MoS$_{2}$ with a carrier concentration of $1.3 \times 10^{13}$ /cm$^{2}$ are only 0.05 and 0.01 for hole and electron doping, respectively. Such a low value of zT is not useful for practical application purposes. With the application of pressure, the zT value under hole and electron doping is found to increase significantly. For holes the zT value is found to increase monotonically with pressure, whereas for electrons a dramatic enhancement is observed starting from 10 GPa. At a pressure of 25 GPa, the room temperature zT values of ML-MoS$_{2}$ are found to be 0.17 and 0.35 for hole and electron doping, which are significantly higher compared to the zero pressure values. Apart from the reduction in lattice thermal conductivity, the gradual increase in hole mobility and relaxation time with pressure act behind the enhancement of the zT value under hole doping. However, for electrons, the increase in thermopower (S) due to the valley convergence in the conduction band (CB) and the involvement of the CB-$\Lambda$ in electronic transport with significantly high carrier mobility acts in unison to result in the remarkable increase in zT. In addition to the pressure induced enhancement, the zT values are found to increase significantly with increasing temperature, as can be seen in Fig. \ref{Fig.8} (a). The highest zT values at 900 K obtained with hole and electron doping are 1.21 and 1.63, respectively. These zT values, obtained with ML-MoS$_{2}$ are comparable to any good commercial thermoelectric material. Also, it is worth noting that the enhanced thermoelectric performance is not only achieved at a specific doping concentration of $1.3 \times 10^{13}$ /cm$^{2}$; instead, the enhancement can be achieved for a range of doping concentrations. To further support this point, the variation in the zT of ML-MoS$_{2}$ at 300 K as a function of doping concentration under different pressures is shown in Fig. \ref{Fig.8} (b). It is evident that a much improved thermoelectric performance can be obtained at the optimal doping concentration when a certain amount of pressure is applied. Notably, the concurrent increase in zT for both electron and hole doping is highly beneficial for its integration in thermoelectric devices, where both the p- and n-legs are preferably made from a single material with high thermoelectric efficiency.

\section{Conclusions}
In summary, first principles calculations have been performed to analyze the hydrostatic strain induced changes in the electronic structure and transport properties of ML-MoS$_2$. The application of external pressure in ML-MoS$_2$ induces a direct to indirect band gap transition at around 10 GPa due to the shifting of the CBM from K to $\Lambda$, which also increases the valley degeneracy with the band edge at the K point still lying close in energy to the CBM at $\Lambda$. Owing to this increase in valley degeneracy, the thermopower (S) and thereby, the power factor (S$^{2}\sigma$) show an increase for the electron doping. Also, the application of hydrostatic pressure results in an increase in electron mobility ($\mu$) and a dramatically enhanced relaxation time ($\tau$), which follows from the low deformation potential associated with CB-$\Lambda$. On the other hand, the lattice thermal conductivity ($\kappa_{\text{L}}$) is also found to decrease significantly with increasing pressure due to the increase in anharmonicity and the resulting phonon scattering. The pressure induced increase in S and $\mu$, and the decrease in $\kappa_{\text{L}}$ act in unison to result in a large increase in zT. At 900 K with an external pressure of 25 GPa, zT values of 1.21 and 1.63 are achieved for the hole and the electron doping, respectively, which are significantly higher compared to that for the zero pressure case. This study, therefore, highlights the importance of hydrostatic pressure in improving the thermoelectric properties of ML-MoS$_2$. Given that the application of hydrostatic pressure increase the efficiency for both the n-type and p-type doping, this fairly straight forward method can be beneficial  for the designing of thermoelectric module, where both legs should be preferably made of the same material. Also, the proposed approach is expected to be equally useful for all other semiconducting TMDCs with similar crystal structures.

\begin{acknowledgments}
	The first-principles calculations have been performed using the supercomputing facility of IIT Kharagpur established under the National Supercomputing Mission (NSM), Government of India and supported by the Centre for Development of Advanced Computing (CDAC), Pune. AB acknowledges SERB POWER grant (SPG/2021/003874) and BRNS regular grant (BRNS/37098) for the financial assistance. SC acknowledges MHRD, India, for financial support.
\end{acknowledgments}

\bibliographystyle{achemso}
\bibliography{biblio}

\end{document}